\documentclass[12pt]{article}
\usepackage{latexsym}
\usepackage{amssymb}
\usepackage{graphicx}
\usepackage{multibox}
\usepackage{amsmath,amsfonts}

\topskip \textwidth 480pt

\def\*{\ast}

\def\ve{\varepsilon}

\def\be{\begin{equation}}
\def\ee{\end{equation}}
\def\bqn{\begin{eqnarray}}
\def\eqn{\end{eqnarray}}

\def\theequation{\thesection.\arabic{equation}}

\newsavebox{\ver}
\newsavebox{\verp}

\newsavebox{\gorp}
\newsavebox{\toch}

\newcommand{\bee}{\begin{eqnarray}}
\newcommand{\eee}{\end{eqnarray}}

\newcommand{\ups}{\upsilon}

\date{}
\begin{document}
\begin{titlepage}
\title{
\begin{flushright}
{\small MIFPA-01-41}\\
~\\
~\\
\end{flushright}
{\bf Evidence for the classical integrability of the complete $AdS_4\times CP^3$ superstring}
~\\
\medskip
\medskip
\medskip
\medskip
\author{Dmitri~Sorokin\footnote{\tt dmitri.sorokin@pd.infn.it}$\,$ and Linus Wulff${~}^*$\footnote{\tt linus@physics.tamu.edu}
~\\
~\\
~\\
{$^*$\it Istituto Nazionale di Fisica Nucleare, Sezione di Padova,}
~\\
{\it via F. Marzolo 8, 35131 Padova, Italia}
~\\
~\\
{$^\dagger$\it George and Cynthia Woods Mitchell Institute}
~\\
{\it for Fundamental Physics and Astronomy,}
~\\
{\it Texas A\&M University, College Station, TX 77843, USA} 
}}
\maketitle



\begin{abstract}
\noindent
 We construct a zero--curvature Lax connection in a sub--sector of the superstring theory on
$AdS_4\times CP^3$ which is not described by the $OSp(6|4)/U(3)\times SO(1,3)$ supercoset
sigma--model. In this sub--sector worldsheet fermions associated to eight broken supersymmetries of
the type IIA background are physical fields. As such, the prescription for the construction of the
Lax connection based on the $Z_4$--automorphism of the isometry superalgebra $OSp(6|4)$ does not do
the job. So, to construct the Lax connection we have used an alternative method which nevertheless
relies on the isometry of the target superspace and kappa--symmetry of the Green--Schwarz
superstring.
\end{abstract}
~

\thispagestyle{empty}
\end{titlepage}
\tableofcontents
\section{Introduction}
The $AdS_4\times CP^3$ background of type IIA superstring theory is not maximally supersymmetric.
It preserves 24 supersymmetries (out of the maximum number of 32) which together with the bosonic
isometries of $AdS_4\times CP^3$ form the supergroup $OSp(6|4)$. It turns out that the type IIA
superspace associated with the $AdS_4\times CP^3$ background which has 32 Grassmann--odd directions
is not a coset superspace of $OSp(6|4)$ \cite{Gomis:2008jt}. So the complete Green--Schwarz
superstring theory on this superspace is not a coset--superspace sigma--model, in contrast
\emph{e.g.} to the maximally supersymmetric type IIB superstring on $AdS_5\times S^5$ described by
the $PSU(2,2|4)/(SO(1,4)\times SO(5))$ sigma--model \cite{Metsaev:1998it}. The worldsheet
$AdS_4\times CP^3$ superstring action can be reduced to an $OSp(6|4)/U(3)\times SO(1,3)$
sigma--model constructed in
\cite{Arutyunov:2008if,Stefanski:2008ik,Fre:2008qc,Bonelli:2008us,D'Auria:2008cw} in those sub--sectors of the classical
configuration space of the theory in which the kappa--symmetry can be used to eliminate eight
fermionic modes of the string associated with the broken supersymmetries. However, this is not always
possible. For instance such a gauge choice is inadmissible when the classical string moves entirely
in $AdS_4$ \cite{Arutyunov:2008if,Gomis:2008jt} or forms a worldsheet instanton wrapping a 2--cycle
inside $CP^3$ \cite{Cagnazzo:2009zh}. In these cases the `broken supersymmetry' fermions are physical
modes, so one should start the analysis of the theory in these sectors from the complete
$AdS_4\times CP^3$ superstring action
\cite{Gomis:2008jt} and, if required, make an alternative choice of the kappa--symmetry gauge (see
e.g. \cite{Grassi:2009yj,Uvarov:2009hf,Uvarov:2009nk,Bykov:2010tv}).

 The classical integrability of the $OSp(6|4)/U(3)\times SO(1,3)$ $\sigma$--model sub--sector of
the theory was demonstrated in \cite{Arutyunov:2008if,Stefanski:2008ik} by constructing a
zero--curvature Lax connection using the same techniques as for the $AdS_5\times S^5$ superstring
\cite{Bena:2003wd}. Such a construction is based on the $Z_4$--automorphism of the isometry
superalgebra and can be applied to any $G/H$ supercoset two--dimensional sigma--model that admits a
$Z_4$--grading. Basically, the prescription is as follows. Take a left--invariant Cartan form
$K^{-1}dK$ (with $K\in G/H$ being a supercoset element) which are used to build the supercoset
sigma--model action
\cite{Metsaev:1998it,Bena:2003wd,Arutyunov:2008if,Stefanski:2008ik}.
The Cartan form takes values in the isometry superalgebra $\mathcal G$ of $G$ and thus can be
expanded in the bosonic generators $M_0$ and $P_2$, and the fermionic generators $Q_1$ and $Q_3$ of
$\mathcal G$
\be\label{cce}
K^{-1}dK=\Omega_0\, M_0+E_2\,P_2+E_1\,Q_1+E_3\,Q_3\,.
\ee
The building blocks of the $G/H$ supercoset sigma--model action are the $G/H$ supervielbeins $E_2$,
$E_1$ and $E_3$, while $\Omega_0$ is the $H$--valued spin connection on $G/H$.

The bosonic generators $M_0$ of the stability subgroup $H$ have zero grading under the
$Z_4$--automorphism and the bosonic coset--space translation generators $P_2$ carry grading two.
The fermionic generators $Q_1$ and $Q_3$ have the $Z_4$--grading one and three, respectively. In
terms of these generators the superalgebra $\mathcal G$ has the following schematic $Z_4$--grading
structure
\bee\label{susyal}
&[M_0,M_0]\sim M_0,\quad [M_0,P_2]\sim P_2,\quad [P_2,P_2]\sim M_0,&\nonumber\\
&
 [M_0,Q_1]\sim Q_1,\quad [M_0,Q_3]\sim Q_3,\quad
[P_2,Q_1]\sim Q_3,\quad [P_2,Q_3]\sim Q_1,&\\
&
 \{Q_1,Q_1\}\sim P_2,\quad
\{Q_3,Q_3\}\sim P_2,\quad \{Q_1,Q_3\}\sim M_0.
&\nonumber
\eee
In the case of the $AdS_4\times CP^3$ superstring $M_0\in so(1,3)\times u(3)$,
$P_2\in \frac{so(2,3)\times su(4)}{so(1,3)\times u(3)}$ and $Q_1$ and $Q_3$ are the 24 fermionic
generators of $OSp(6|4)$, see Appendix \ref{ospalgebra}.

The worldsheet Lax connection one--form which takes values in $\mathcal G$ is constructed by taking
the sum of the components of the Cartan form \eqref{cce} and their worldsheet Hodge--duals with
some arbitrary coefficients, namely
\be\label{L}
L=\Omega_0\,M_0+(l_1 E_2+l_2\ast E_2)\,P_2+l_3\,E_1\,Q_1+l_4\,E_3Q_3.
\ee
Then one imposes the requirement that the curvature associated with the connection $L$ vanishes
\be\label{ZC}
dL-L\wedge L=0
\ee
(the exterior derivative acts from the right, and in what follows we shall not explicitly write
the wedge--product). The sigma--model equations of motion and the $Z_4$--grading structure of the
superalgebra
\eqref{susyal} ensure that the coefficients in the definition of the zero--curvature Lax connection
\eqref{L} are expressed in terms of a single independent \emph{spectral} parameter, \emph{e.g.}
$l_1=\frac{1+z^2}{1-z^2}$.

By performing a gauge transformation of
\eqref{L} one can get another form of the Lax connection \cite{Bena:2003wd} associated with
right--invariant Cartan forms $dKK^{-1}$
\be\label{mL}
{\mathcal L}=KLK^{-1}-dKK^{-1}\,, \qquad d\mathcal L-\mathcal L\, \mathcal L=0\,.
\ee
Having at hand the Lax connection, one can then derive an infinite set of conserved charges of the
integrable model from the holonomy of the Lax connection by constructing a corresponding monodromy
matrix and the algebraic curve (see
\emph{e.g.}
\cite{Bena:2003wd,Zarembo:2010yz} for more details and references therein).

In the case of the complete Green--Schwarz theory (\emph{i.e.} when the kappa--symmetry is not
fixed at all) the superstring moves in $AdS_4\times CP^3$ superspace with thirty two Grassmann--odd
directions and the eight worldsheet fermionic fields associated to the broken supersymmetry
contribute to the structure of the supervielbeins $E_2$, $E_1$ and $E_3$  and to the connection
$\Omega_0$ thus spoiling their nature as the $G/H$ Cartan forms. As a result, as one can check by
direct calculations, the $OSp(6|4)$ Lax connection of the form
\eqref{L} or \eqref{mL} constructed from $\Omega_0$, $E_2$, $E_1$ and $E_3$ which include the dependence on these eight
fermions will not have zero curvature for any non--trivial choice of the coefficients. Therefore, a
modification of the form of
\eqref{L} or \eqref{mL} by additional terms depending on the extra eight fermions is required for
restoring the zero curvature condition \eqref{ZC}. The goal of this paper is to reveal the
structure of these terms.

To construct the Lax connection which includes broken supersymmetry fermions we have found helpful
to look at the form of conserved Noether currents associated with the $OSp(6|4)$ isometry. In this
respect it is more convenient to consider the Lax connection in the form
\eqref{mL} which, in a certain sense, has closer relation to a $G/H$ sigma--model conserved current
having the form \cite{Bena:2003wd,Arutyunov:2008if}
\be\label{ghc}
J_{coset}=K\left(E_2\,P_2+\frac{1}{2}\,*(E_1Q_1-E_3Q_3)\right)K^{-1}\,.
\ee
The paper is organized as follows. In Section \ref{action} we consider the $AdS_4\times CP^3$
superstring action truncated to the second order in fermions and show that there exist different
forms of the Lax connection, related to each other by local $OSp(6|4)$ transformations, which have
zero curvature at least to the second order. When the eight broken supersymmetry fermions are put
to zero the Lax connection reduces (modulo a gauge transformation) to the supercoset Lax connection
of
\cite{Arutyunov:2008if,Stefanski:2008ik}. The reconstruction of higher order fermionic terms in the
Lax connection becomes technically more and more complicated with each order and we have not been
able to accomplish the construction in the complete theory with 32 fermions. So in Section 3 we
consider a simpler sub--sector of the theory in which the superstring moves only in an $AdS_4$
superspace with eight fermionic directions associated with broken supersymmetries. This sub--sector
of the theory is not reachable by the $OSp(6|4)$ supercoset sigma--model and can be regarded as a
model of an $\mathcal N=2$, $D=4$ superstring in the $AdS_4$ background with completely broken
supersymmetries
\cite{Gomis:2008jt}. Nevertheless, this model is invariant under the four--parameter
kappa--symmetry, in addition to the purely bosonic isometry $SO(2,3)$ of $AdS_4$ and $SO(2)$
transformations of the two Majorana fermions. So, surprisingly, the integrability of its fermionic
sector is not related to target space supersymmetry. To simplify the construction of the full Lax
connection in this model, in Section 4 we gauge fix kappa--symmetry and perform worldsheet
T--duality transformations along the $AdS_4$ Minkowski boundary following the results of
\cite{Grassi:2009yj}. In Subsection \ref{D4L} we give the explicit form of the kappa--symmetry gauge--fixed Lax
connection of the $AdS_4$ superstring to all orders in fermions thus giving more evidence for the
classical integrability of the complete $AdS_4\times CP^3$ superstring itself. Section \ref{cd} is
devoted to a summary of the obtained results and discussion of the possibility of their
generalization and application to strings in other supergravity backgrounds. Our notation and
conventions are given in Appendices A and B, and in Appendices C and D we have collected various
formulas and relations which have been used to construct the Lax connections.

\section{$AdS_4\times CP^3$ superstring in the quadratic approximation in fermions}\label{action}
\setcounter{equation}0

\subsection{The action and equations of motion}
We first check that a zero--curvature Lax connection does exist in the complete $AdS_4\times CP^3$
superstring theory at least up to the second order in the fermionic fields. To this end we start
with the $AdS_4\times CP^3$ superstring action truncated to the second order in fermions as in
\cite{Cvetic:1999zs}. In the notation and conventions of \cite{Cagnazzo:2009zh} the action has the
following form
\bee\label{action2}
S&=&-\frac{e^{\frac{2}{3}\phi_0}}{4\pi\alpha'}\,\int d^2\xi\, \sqrt {-h}\, h^{IJ}\, {e}_{I}{}^{A}
{e}_{J}{}^{B} \eta_{AB}\nonumber\\
&-&
\frac{e^{\frac{2}{3}\phi_0}}{2\pi\alpha'}\,\int
d^2\xi\,\Theta(\sqrt{-h}\,h^{IJ}-\varepsilon^{IJ}\Gamma_{11})\big[i\,e_I{}^A\Gamma_A\nabla_J\Theta
-\frac{1}{R}e_I{}^Ae_J{}^B\Gamma_A\mathcal P_{24}\gamma^5\Gamma_B\Theta\big]
\eee
where $h_{IJ}(\xi)$ ($I,J=0,1$) is the intrinsic (auxiliary) worldsheet metric,
$e_I{}^A=\partial_IX^Me_M{}^A(X)$ are the worldsheet pullbacks of the $AdS_4\times CP^3$ vielbeins
($M=0,1,\cdots,9$ are the $D=10$ space-time indices and $A=0,1,\cdots,9$ are the tangent space
indices). $X^M=(x^{\hat m},y^{m'})$ are $AdS_4\times CP^3$ coordinates ($\hat m=0,1,2,3;$
$m'=1',\cdots 6'$), $\nabla\Theta=(d-\frac{1}{4}\,\omega^{AB}\,\Gamma_{AB})\Theta$ is the
worldsheet pullback of the conventional $AdS_4\times CP^3$ covariant derivative and $\mathcal
P_{24}$ is the projector which splits the 32 fermionic coordinates $\Theta^{\underline\alpha}$
($\underline\alpha=1,\cdots,32$) into 24 fermionic coordinates $\vartheta$ corresponding to the 24
unbroken supersymmetries of the $AdS_4\times CP^3$ background and 8 `broken supersymmetry'
coordinates $\upsilon$
\be\label{p6}
{\mathcal P}_{24}={1\over 8}(6+iJ_{a'b'}\,\Gamma^{a'b'}\,\gamma^7)\,,\qquad \vartheta
\equiv{\mathcal P}_{24}\,\Theta\,,\qquad \upsilon\equiv (1-{\mathcal P}_{24})\,\Theta.
\ee
In \eqref{p6} $J_{a'b'}=-J_{b'a'}$ is the K\"ahler form on $CP^3$, $\Gamma^{a'}$ are $D=10$
Dirac matrices along the six $CP^3$ directions ($a'=1',\cdots,6'$) and $\gamma^7=i\Gamma^{1'}\cdots\Gamma^{6'}$ is the product of all of them. The presence in the action
\eqref{action2} of the projector $\mathcal P_{24}$ is due to the interaction of the string with
the constant Ramond--Ramond $F_4\sim dx^{0}dx^1dx^2dx^3$ and $F_2\sim dy^{a'}dy^{b'}J_{a'b'}$
fluxes of type IIA supergravity on $AdS_4\times CP^3$. $\gamma^5=i\Gamma^{0123}$ is the product of
the four gamma--matrices with $AdS_4$ indices. Finally, $\phi_0$ is the vacuum expectation value of
the dilaton and $R$ is related to the $CP^3$ radius in the string frame
$R_{CP^3}=e^{\frac{\phi_0}{3}}\,R$. See Appendix A for more details of our notation and
conventions.

The bosonic field equations which follow from \eqref{action2} are
\bee\label{bosoneq}
&\nabla_I\left[\sqrt{-h}\,h^{IJ}e_J{}^A+i\Theta(\sqrt{-h}\,h^{IJ}
-\varepsilon^{IJ}\Gamma_{11})\big(\Gamma^A\nabla_J\Theta
+\frac{2i}{R}\,e_J{}^B\Gamma^A\mathcal P_{24}\gamma^5\Gamma_B\Theta\big)\right]&\nonumber\\
&&\\
 &-\frac{i}{4}\Theta\,(\sqrt{-h}\,h^{IJ}-\varepsilon^{IJ}\Gamma_{11})\Gamma_D{}^{BC}\Theta
\,R_{BCE}{}^{A}\,e_I{}^De_J{}^E=0&\nonumber
\eee
where $R_{BCE}{}^{A}$ is the curvature of $AdS_4\times CP^3$ (see Appendix A).

The Virasoro constraints are
\bee\label{Virasoro}
&{e}_{I}{}^{A} {e}_{J}{}^{B}
\eta_{AB}-
2i\Theta\big(e_{(I}{}^A\Gamma_A\nabla_{J)}\Theta
+\frac{i}{R}e_{(I}{}^Ae_{J)}{}^B\Gamma_A\mathcal P_{24}\gamma^5\Gamma_B\Theta\big)\nonumber\\
&&\\
 &=
\frac{1}{2}h_{IJ}\,h^{KL}\,\left[e_{K}{}^{A}\,{e}_{L}{}^{B}
\eta_{AB}-
2i\Theta\big(e_K{}^A\Gamma_A\nabla_L\Theta +\frac{i}{R}e_K{}^Ae_L{}^B\Gamma_A\mathcal
P_{24}\gamma^5\Gamma_B\Theta\big)\right]\,,\nonumber
\eee
where the round brackets embracing the indices denote symmetrization
$X_{(I}Y_{J)}=\frac{1}{2}\,(X_{I}Y_{J}+X_{J}Y_{I})$.

The fermionic equations  are
\be\label{fermilinear1}
(\sqrt{-h}\,h^{IJ}-\varepsilon^{IJ}\Gamma_{11})\big(e_I{}^A\Gamma_A\nabla_J\Theta
+\frac{i}{R}e_I{}^Ae_J{}^B\Gamma_A\mathcal
P_{24}\gamma^5\Gamma_B\Theta\big)-\frac{1}{2}\,\nabla_I(\sqrt{-h}\,h^{IJ}e_J{}^A)\Gamma_A\Theta=0\,.
\ee
In virtue of the bosonic equations \eqref{bosoneq}, the last term in \eqref{fermilinear1} is of the
third order in fermions and can be skipped in the linear approximation.

\subsubsection{Comment on the relation to the supercoset sigma--model}

When the fermionic fields $\upsilon$ are zero the superstring equations of motion reduce to the
bosonic equation
\bee\label{bosoneq1}
&\nabla_I\left[\sqrt{-h}\,h^{IJ}e_J{}^A+i\vartheta(\sqrt{-h}\,h^{IJ}
-\varepsilon^{IJ}\Gamma_{11})\big(\Gamma^A\nabla_J\vartheta
+\frac{2i}{R}\,e_J{}^B\Gamma^A\mathcal P_{24}\gamma^5\Gamma_B\vartheta\big)\right]&\nonumber\\
 &-\frac{i}{4}\vartheta\,(\sqrt{-h}\,h^{IJ}-\varepsilon^{IJ}\Gamma_{11})\Gamma_D{}^{BC}\vartheta
\,R_{BCE}{}^{A}\,e_I{}^De_J{}^E=0&
\eee
and the fermionic equations
\bee\label{fermilinear12}
&(\sqrt{-h}\,h^{IJ}-\varepsilon^{IJ}\Gamma_{11})\,e_I{}^A\mathcal P_{24}\Gamma_A\mathcal
P_{24}\big(\nabla_J\vartheta +\frac{i}{R}e_J{}^B\gamma^5\Gamma_B\vartheta\big)=0\,,\\
&&\nonumber\\
 &(\sqrt{-h}\,h^{IJ}-\varepsilon^{IJ}\Gamma_{11})\,e_I{}^{A}(1-\mathcal
P_{24})\Gamma_{A}\mathcal P_{24}\big(\nabla_J\vartheta
+\frac{i}{R}e_J{}^B\gamma^5\Gamma_B\vartheta\big)=0. &
\label{fermilinear13}\eee
Eqs. \eqref{bosoneq1} and \eqref{fermilinear12} are the equations of motion of the $OSp(6|4)$
supercoset sigma--model in the quadratic approximation in fermions. However, the complete
Green--Schwarz superstring action gives one more fermionic equation of motion
 which (when $\upsilon=0$) produces an additional equation for the 24
fermions $\vartheta$ \eqref{fermilinear13}. This eight--component equation does not directly follow
from the supercoset action, but it should not be independent of
\eqref{fermilinear12} and just manifests the fact that, when the partial kappa--symmetry gauge
$\upsilon=0$ is admissible, the residual kappa--symmetry of the supercoset model has eight
independent components, such that the number of physical fermionic modes of $\vartheta$ is sixteen.

To show that the fermionic equations \eqref{fermilinear12} and \eqref{fermilinear13} are linearly
dependent let us rewrite them in an equivalent form as follows
\bee\label{fermilinear12E}
&\mathcal P_{24}(1-\Gamma)\,h^{IJ}e_I{}^A\Gamma_A\mathcal
P_{24}\big(\nabla_J\vartheta +\frac{i}{R}e_J{}^B\gamma^5\Gamma_B\vartheta\big)=0\,,\\
&&\nonumber\\
 &(1-\mathcal
P_{24})(1-\Gamma)\,h^{IJ}\,e_I{}^{A}\Gamma_{A}\mathcal P_{24}\big(\nabla_J\vartheta
+\frac{i}{R}e_J{}^B\gamma^5\Gamma_B\vartheta\big)=0\,, &
\label{fermilinear13E}\eee
where $\Gamma=\frac{1}{2\sqrt{-h}}\varepsilon^{IJ}e^A_I\,e^B_{J}\,\Gamma_{AB}\Gamma_{11}$,
$(\Gamma)^2=1$ and $\frac{1}{2}(1-\Gamma)$ is the canonical kappa--symmetry projector of the type
IIA superstring. The two equations can, therefore, be combined into
\be\label{combined}
(1-\Gamma)\,h^{IJ}e_I{}^A\Gamma_A\mathcal P_{24}\big(\nabla_J\vartheta
+\frac{i}{R}e_J{}^B\gamma^5\Gamma_B\vartheta\big)=0\,.
\ee
We shall now show that eq. \eqref{combined} actually follows from eq. \eqref{fermilinear12E}. To
this end let us note that in the sector of classical string solutions in which the kappa--symmetry
gauge $ \upsilon=0$ is admissible, the projectors $\mathcal P_{24}$ and $\frac{1}{2}(1\pm\Gamma)$
do not commute \cite{Gomis:2008jt}, their commutator $[\Gamma,\mathcal P_{24}]$ being a non
degenerate matrix. Therefore, multiplying eq. \eqref{fermilinear12E} by $(1+\Gamma)$ we have
\be\label{combined1}
[\Gamma,\mathcal P_{24}]\,(1-\Gamma)\,h^{IJ}e_I{}^A\Gamma_A\mathcal P_{24}\big(\nabla_J\vartheta
+\frac{i}{R}e_J{}^B\gamma^5\Gamma_B\vartheta\big)=0\,.
\ee
Since $[\Gamma,\mathcal P_{24}]$ is invertible we can multiply the above equation by the inverse of
$[\Gamma,\mathcal P_{24}]$ and get eq. \eqref{combined} from which the equation
\eqref{fermilinear13E} follows.

On the other hand, in the sub--sector in which the classical string moves in $AdS_4$ only (i.e. the
$CP^3$ embedding coordinates $y^{m'}$ are constants), this kappa--gauge is not admissible
($[\Gamma,\mathcal P_{24}]=0$) and putting $\upsilon$ to zero results in loosing four physical
fermionic modes associated with $\upsilon$ \cite{Arutyunov:2008if,Gomis:2008jt}. This can be seen
from the structure of the fermionic equations \eqref{fermilinear12} and \eqref{fermilinear13} (or
\eqref{fermilinear12E} and \eqref{fermilinear13E}). Since $y^{m'}$ are constants  and if $\upsilon$ is set to zero, eq.
\eqref{fermilinear13} (or \eqref{fermilinear13E})
vanishes identically and one is left with eq.
\eqref{fermilinear12} (or \eqref{fermilinear12E}) which, since the projector $\mathcal P_{24}$ commutes with the $\Gamma^{\hat a}$ along the
$AdS_4$ directions, reduces to the fermionic equation in $AdS_4$
\be\label{fermilinear1222}
(1-\Gamma)\,h^{IJ}\,e_I{}^{\hat a}(x)\Gamma_{\hat a}\big(\nabla_J\vartheta +\frac{i}{R}e_J{}^{\hat
b}\gamma^5\Gamma_{\hat b}\vartheta\big)=0\,.
\ee
where now $\Gamma=\frac{1}{2\sqrt{-h}}\,\varepsilon^{IJ}e_I{}^{\hat a}e_J{}^{\hat b}\,\Gamma_{\hat
a\hat b}\Gamma_{11}$, $(\Gamma)^2=1$. The projector $\frac{1}{2}(1-\Gamma)$ (which now commutes
with $\mathcal P_{24}$) implies that among 24 equations \eqref{fermilinear1222} only 12 are
independent. Hence $\vartheta$ contain only 12 physical modes while the total number must be
sixteen. The missing four physical fermions are half of $\upsilon$ which were put to zero `by
hand', while another half of $\upsilon$ can be gauged away by kappa--symmetry.

\subsection{Noether currents}

Under the $OSp(6|4)$ isometries the Type IIA superspace coordinates $X^M$ and $\Theta$ transform as
follows (up to the second order in fermions)
\bee\label{isometry}
&\delta X^M\,e_M{}^A(X)=K^A(X)+i\Theta\Gamma^A\,\Xi(X),& \nonumber\\
&&\nonumber\\
 &\delta \vartheta=\mathcal
P_{24}\delta\Theta=\Xi(X)+\frac{1}{4}(K^M\,\omega_M{}^{AB}(X)-\nabla^A\,K^B)\,\mathcal
P_{24}\Gamma_{AB}\mathcal
P_{24}\Theta\,,&\\
&&\nonumber\\
&\delta\upsilon=(1-\mathcal
P_{24})\,\delta\Theta=\frac{1}{4}(K^M\,\omega_M{}^{AB}(X)-\nabla^A\,K^B)\,(1-\mathcal
P_{24})\,\Gamma_{AB}\,(1-\mathcal P_{24})\,\Theta, &\nonumber
\eee
where $K^A(X)=K^M(X)\,e_M{}^A(X)$ are the $AdS_4\times CP^3$ Killing vectors. More precisely,
$K^A(X)$ are the Killing vectors $K^A_{\mathcal I}(X)$ contracted with constant $SO(2,3)\times
SU(4)$ transformation parameters $\Lambda^{\mathcal I}$,\emph{ i.e.} $K^A(X)=K^A_{\mathcal
I}(X)\,\Lambda^{\mathcal I}\,,$ where $\mathcal I$ is associated with the 25 generators of the
$SO(2,3)\times SU(4)$ isometries. Note that, like the spin connection $\omega^{AB}$,
$\nabla^AK^B=-\nabla^BK^A$ takes values in the stability subalgebra $so(1,3)\times u(3)$ of the
$AdS_4\times CP^3$ isometry. Properties of the Killing vectors of symmetric spaces $G/H$ are given
in Appendix D \ref{AD}.

\noindent
$\Xi$ are 24 supersymmetry parameters of $OSp(6|4)$ satisfying the $AdS_4\times CP^3$ Killing
spinor equation
\begin{equation}\label{KS}
\nabla\Xi+\frac{i}{R}e^A\,\mathcal
P_{24}\gamma^5\Gamma_A\Xi=0\,, \qquad
\Xi^{\underline\alpha}(X)=\epsilon^{\underline\mu}\,\,\Xi_{\underline\mu}{}^{\underline\alpha}(X),\qquad
\Xi\equiv\mathcal P_{24}\,\Xi(X)\,,
\end{equation}
$\Xi_{\underline\mu}{}^{\underline\alpha}(X)$ are $AdS_4\times CP^3$ Killing spinors and
$\epsilon^{\underline\mu}=(\mathcal P_{24}\epsilon)^{\underline\mu}$ are 24 constant Grassmann--odd
parameters.

Note that the terms in the variation of the fermions which are proportional to $\Gamma_{AB}$ are
the compensating $SO(1,3)\times U(3)$ stability group transformations induced by the isometries in
the (co)tangent space of $AdS_4\times CP^3$. Note also that in the linear order in fermions the
eight spinor fields $\upsilon$ are not transformed by supersymmetry. The action of the isometry
group $OSp(6|4)$ on these fermions is such that it takes the form of induced $SO(1,3)\times U(1)$
rotations with parameters depending on $X$, $\vartheta$ and the $OSp(6|4)$ parameters
\be\label{deltav}
\delta \upsilon
=\frac{1}{4}\,\Lambda_{AB}(\epsilon,X,\vartheta)\,\Gamma^{AB}\,\upsilon\,.
\ee
Therefore, the first nontrivial term in the supersymmetry variation of $\upsilon$ is quadratic in
fermionic fields.

To avoid possible confusion, let us note that in the expressions for the conserved currents and in
the Lax connections considered below, $K^A(X)$ and $\Xi(X)$ stand for the Killing vectors and
spinors contracted with the corresponding bosonic and fermionic generators of the $OSp(6|4)$
isometry (see Appendix \ref{ospalgebra}) and not with constant parameters like in eqs. \eqref{isometry}
and \eqref{KS}.

The following relations between the Killing vectors and spinors contracted with the $OSp(6|4)$
generators reflect the structure of the $OSp(6|4)$ superalgebra \eqref{sp411}--\eqref{osp64}
\bee\label{killing}
&K_A(X)\doteq k(X)P_Ak^{-1}(X)\,,\qquad\gamma^5\Xi(X)\doteq
k(X)Qk^{-1}(X)\,,&\\
&\nabla_AK_B\doteq-\frac{1}{2}R_{AB}{}^{CD}\,k(X)M_{CD}k^{-1}(X)\,,&\nonumber
\eee
where $k(X)$ is an $\frac{SO(2,3)\times SU(4)}{SO(1,3)\times U(3)}$ coset element  of the bosonic
isometry, and
\begin{eqnarray}\label{Kvs}
&[K_A,\Xi]=-\frac{i}{R}\Xi\Gamma_A\gamma^5\mathcal P_{24}\,,&\nonumber\\
&{}[\nabla_AK_B,\Xi]=-\frac{1}{4}R_{AB}{}^{CD}\Xi\Gamma_{CD}\mathcal P_{24}\,,&\\
&\{\Xi,\Xi\}=2i\,\mathcal P_{24}\gamma^5\Gamma^A\gamma^5\mathcal P_{24}\,K_A-\frac{R}{2}\mathcal
P_{24}\Gamma^{AB}\gamma^5\mathcal P_{24}\, \nabla_AK_B\,.&\nonumber
\end{eqnarray}

The conserved Noether current associated with the $SO(2,3) \times SU(4)$ invariance of the action
\eqref{action2} is
\bee\label{JB}
J^I_B&=&\sqrt{-h}\,h^{IJ}\,e_J{}^A\,K_A
+i\Theta(\sqrt{-h}\,h^{IJ}-\varepsilon^{IJ}\Gamma_{11})\,\big[\Gamma^A\nabla_J\Theta+
\frac{2i}{R}e_J{}^B\Gamma^A\mathcal P_{24}\gamma^5\Gamma_B\Theta\big]\,K_A\nonumber\\
&&\\
&&-\frac{i}{4}\,\Theta(\sqrt{-h}\,h^{IJ}+\varepsilon^{IJ}\Gamma_{11})\,e_J{}^A\,\Gamma_A{}^{BC}\Theta\,\nabla_B\,K_C\,\nonumber
\eee
and the conserved (fermionic) supersymmetry current (up to the leading order in fermions) is
\bee\label{Jsusy}
J_{F}^I&=&\frac{i}{2R}\Big(\sqrt{-h}\,h^{IJ}\,e_J{}^A\,\Theta\Gamma_A\Xi
+\Theta(\sqrt{-h}\,h^{IJ}+2\varepsilon^{IJ}\Gamma_{11})\,e_J{}^A\,\Gamma_A\Xi(X)\Big)\,\nonumber\\
&=&\frac{i}{R}\,\Theta(\sqrt{-h}\,h^{IJ}+\varepsilon^{IJ}\Gamma_{11})\,e_J{}^A\,\Gamma_A\Xi(X)\,,
\eee
where the factor of 2 in the last term of the first line appears because the action is invariant
under supersymmetry only up to a boundary term which must therefore be subtracted from the current
to make it conserved. The currents are normalized to be dimensionless (the dimensions of $\Xi$ and
$K_A$ are $1/\sqrt{R}$ and $1/R$ respectively). The sum of $J_B$ and $J_F$ is the conserved current
taking values in the $OSp(6|4)$ superalgebra
\be\label{Jtotal}
J=J_B+J_F.
\ee
Let us now compare this current with the conserved current of the $OSp(6|4)$ supercoset
sigma--model which describes the string with $\upsilon=(1-\mathcal P_{24})\Theta=0$. As we have
mentioned in the Introduction, the supercoset model conserved current has the following form (in
our conventions)
\begin{equation}\label{Jv0}
J_{coset}(X,\vartheta)=K(X,\vartheta)\,\Lambda\,K^{-1}(X,\vartheta)\doteq
K(X,\vartheta)\,\Big(E^AP_A+\frac{1}{2}Q\Gamma_{11}\ast E\Big)\,K^{-1}(X,\vartheta)\,,
\end{equation}
where $E^A(X,\vartheta)$ and $E^{\underline\alpha}(X,\vartheta)$ are components of the
$OSp(6|4)$--valued Cartan form
\begin{equation}\label{Cartan}
K^{-1}dK(X,\vartheta)=  E^AP_A+
E^{\underline\alpha}Q_{\underline\alpha}+\frac{1}{2}\Omega^{AB}M_{AB},
\end{equation}
$\Omega^{AB}(X,\vartheta)$ is the spin connection on $\frac{OSp(6|4)}{SO(1,3)\times U(3)}$ and
$K(X,\vartheta)$ is a coset representative. Up to the second order in fermions the supervielbeins
and spin connection of the $OSp(6|4)$ supercoset are given by
\bee\label{EO}
E^A&=&e^A(X)+i\vartheta\Gamma^A\,E\,,\\
E^{\underline\alpha}&=&\nabla\vartheta^{\underline\alpha} +\frac{i}{R}e^B\,(\mathcal P_{24}\gamma^5\Gamma_B\,\vartheta)^{\underline\alpha}\,,\nonumber\\
\Omega^{AB}&=&\omega^{AB}(X)-\frac{2}{R}\vartheta\Gamma^{[A}\mathcal{P}_{24}\gamma_5\Gamma^{B]}E\,.\nonumber
\eee
The current \eqref{Jv0} is conserved $(d*J_{coset}=0)$ as a consequence of the sigma--model equations of
motion
\be\label{sigmaeof}
d\ast\Lambda-[K^{-1}dK,\ast\Lambda]=0.
\ee
Using a supercoset element of the form $K=k(X)\,e^{\vartheta Q}$, the $OSp(6|4)$ superalgebra
(Appendix A.4) and eqs.
\eqref{killing} we then have
\begin{eqnarray}\label{J=J}
J_{coset}&=&
  E^AK_A
+e^Ak\vartheta[Q,P_A]k^{-1} +\frac{1}{2}e^Ak[\vartheta Q,[\vartheta Q,P_A]]k^{-1}
+\frac{1}{2}kQk^{-1}\Gamma_{11}\ast  E
\nonumber\\
&&{} +\frac{1}{2}k\vartheta\{Q,Q\}\Gamma_{11}\ast  E k^{-1}
\nonumber\\
&=& J|_{\upsilon=0} -\frac{R}{8}\ast d(\vartheta\Gamma^{AB}\gamma^7\vartheta\,\nabla_AK_B) -\frac{1}{2}\ast
d(\Xi\gamma^7\vartheta)\,,
\end{eqnarray}
where $J=J_B+J_F$ \eqref{Jtotal} is the Noether current directly derived from the quadratic
Green-Schwarz action. The two conserved currents therefore differ only by total derivative terms,
as should be the case. A useful relation in checking eq. \eqref{J=J} is
\begin{equation}
\mathcal{P}_{24}\Gamma_{[A}\mathcal{P}_{24}\gamma^5\Gamma_{B]}\mathcal{P}_{24}
=-\frac{R^2}{8}R_{AB}{}^{CD}\mathcal{P}_{24}\Gamma_{CD}\gamma^5\mathcal{P}_{24}\,.
\end{equation}

\subsection{Lax connections to the second order in fermions}\label{laxquad}
\subsubsection{The supercoset sigma--model Lax connection}
The Lax connection \eqref{mL} of the $OSp(6|4)$ supercoset sigma--model can be written in the
following form in terms of the conserved current \eqref{Jv0} and components of the Cartan form
\eqref{Cartan}
\bee\label{Lsigma1}
\mathcal L_{coset}
&=& K\Big(\alpha_1 E^AP_A+\alpha_2\ast E^AP_A+\beta_1Q\Gamma_{11}E+(1+\beta_2)QE\Big)K^{-1}
\nonumber\\
&=&K\Big(\alpha_1\,E^A\,P_A
+(1+\beta_2)\,QE+(\beta_1-\frac{\alpha_2}{2})\,Q\Gamma_{11}E\Big)K^{-1}+\alpha_2\,\ast J_{coset}\,,
\eee
where
\begin{eqnarray}\label{coef}
\alpha_1&=&\frac{2z^2}{1-z^2}\,,\nonumber\\
\alpha_2^2&=&\alpha_1^2+2\alpha_1\,,\nonumber\\
\beta_1&=&\mp\sqrt{\frac{\alpha_1}{2}}\,,\nonumber\\
\beta_2&=&\pm\frac{\alpha_2}{\sqrt{2\alpha_1}}\,.
\end{eqnarray}
The specific dependence of the coefficients on the spectral parameter $z$ ensures the zero
curvature of the Lax connection\footnote{The numerical coefficients in eq.
\eqref{Lsigma1} are related to those in eq. \eqref{L} and those of \cite{Arutyunov:2008if} (eq.(4.1) therein) as follows
$\alpha_1=l_1-1$, $\alpha_2=l_2$, $\beta_1=\frac{l_3-l_4}{2}$ and $\beta_2=-\frac{l_3+l_4}{2}$.}
\cite{Bena:2003wd,Arutyunov:2008if,Stefanski:2008ik}.
Note that the $Z_4$--automorphism splitting of the fermionic $OSp(6|4)$ generators $Q$  and the
corresponding fermionic components of the Cartan form is simply made by the $D=10$ chirality
projectors $\frac{1}{2}(1\mp\Gamma_{11})$ (see Appendix A.4).

\subsubsection{Lax connection of the complete $AdS_4\times CP^3$ superstring}
When the extra eight fermionic degrees of freedom $\upsilon$ are switched on, they contribute to
the supervielbeins, superconnection, conserved current and equations of motion and, as a
consequence, the form of the Lax connection should be modified to account for this. In contrast to
the case of the $OSp(6|4)$ supercoset $Z_4$--grading, it is not obvious which is the
group--theoretical structure that would allow one to guess the dependence of $\mathcal L$ on
$\upsilon$. So, to find this dependence we shall use a brute--force method, i.e. we will try to
build the Lax connection out of components of the conserved currents $J_B$ \eqref{JB} and $J_F$
\eqref{Jsusy}, which depend on the extra fermions $\upsilon$, by introducing them with arbitrary
coefficients in the Lax connection. The dependence of these coefficients on the spectral parameter
is then determined by the zero--curvature condition. This procedure is akin to the construction of
Lax connections for two--dimensional supersymmetric non--linear sigma--models considered in
\cite{Curtright:1979am}. The Lax connection constructed in this way has the following form
\be\label{alter}
L=L_B+L_F\,
\ee
where the bosonic isometry part is
\begin{equation}
L_B=\alpha_1e^AK_A+\alpha_2\ast J_B+\alpha_2^2J^{AB}\nabla_AK_B+\alpha_1\alpha_2\ast
J^{AB}\nabla_AK_B\,,
\end{equation}
and the supersymmetry part is
\begin{equation}
L_F=-\alpha_2\beta_1 J_F+\alpha_2\beta_2\ast J_F\,.
\end{equation}
$J^{AB}$ stands for the term in the bosonic isometry current \eqref{JB} which is contracted with
$\nabla_AK_B$, namely $J_B=J^AK_A+J^{AB}\nabla_AK_B$, and $\alpha_1$, $\alpha_2$, $\beta_1$ and $\beta_2$ are the same as
in \eqref{coef}.

It is not very difficult to verify that this Lax connection indeed has zero curvature. To check the zero--curvature condition one should use the conservation of the Noether current, the equations of motion as well as the relations
\bee\label{JAB}
\nabla J^{AB}&=&-e^{[A}(J^{B]}-e^{B]})
-\frac{1}{2R}e^Ce^D\,\Theta\Gamma_C\mathcal{P}_{24}\Gamma^{AB}\gamma^5\mathcal{P}_{24}\Gamma_D\Theta
\nonumber\\
&&+\frac{1}{2R}e^C\ast
e^D\,\Theta\Gamma_C\mathcal{P}_{24}\Gamma^{AB}\gamma^5\mathcal{P}_{24}\Gamma_D\Gamma_{11}\Theta\,,
\nonumber\\
&&\\
dJ_F&=&\frac{i}{R}d(e^A\,\Theta\Gamma_A\Xi-\ast e^A\,\Theta\Gamma_A\Gamma_{11}\Xi)
\nonumber\\
&=&
\frac{2}{R^2}e^Ae^B\,\Theta\Gamma_A\mathcal{P}_{24}\gamma^5\Gamma_B\Xi
-\frac{2}{R^2}e^A\ast e^B\,\Theta\Gamma_A\mathcal{P}_{24}\gamma^5\Gamma_B\Gamma_{11}\Xi\nonumber
\eee
and the symmetry properties of the $\Gamma$--matrices.

Note that the construction of this Lax connection does not make use (at least directly) of the
$Z_4$--grading of the $OSp(6|4)$ superalgebra but only the $Z_2$--grading of its bosonic
subalgebra. Its form is different from the $\upsilon$--fermion extension of the supercoset Lax
connection \eqref{Lsigma1} (\emph{e.g.} the former does not have terms linear in $d\Theta$, while
such terms are present in the latter). We will now show that the two Lax connections are related by
an $OSp(6|4)$ gauge transformation.

\subsubsection{Relation to the supercoset Lax connection}
When $\upsilon=0$ the Lax connection \eqref{alter} constructed above should be related to the
supercoset Lax connection in eq. (\ref{Lsigma1}) by a gauge transformation, so that
\begin{equation}
\mathcal L_{coset}=g^{-1}L|_{\upsilon=0}\,g+g^{-1}dg\,.
\end{equation}
for some $g\in OSp(6|4)$. It is possible to show that this is indeed the case and with a bit of
algebra one finds that the supergroup element
\bee
g(X,\vartheta;\alpha_2,\beta_1,\beta_2)&=&k(X)\,e^{\frac{\alpha_2R}{16}\vartheta\Gamma^{AB}\gamma^7\vartheta\,R_{AB}{}^{CD}M_{CD}}
\,e^{-\beta_1\vartheta\Gamma_{11}Q}\,e^{-(1+\beta_2)\vartheta
Q}\,k^{-1}(X)
\nonumber\\
&=&e^{-\frac{\alpha_2R}{8}\vartheta\Gamma^{AB}\gamma^7\vartheta\,\nabla_AK_B}\,
e^{\beta_1\vartheta\gamma^7\Xi}\,e^{-(1+\beta_2)\vartheta\gamma^5\Xi}
\eee
does the job. If we apply this gauge transformation to the Lax connection $L$ \eqref{alter} without
setting $\upsilon$ to zero we obtain the supercoset Lax connection extended with the terms up to
quadratic order in $\upsilon$
\be\label{glg}
\mathcal L=g^{-1}L\,g+g^{-1}dg\,.
\ee

The Lax connections constructed above have zero curvatures only up to the quadratic order in
fermions. To get zero curvature also at quartic and higher orders in fermions one should add
to the Lax connection \eqref{alter} or \eqref{glg} corresponding higher--order fermionic
$\upsilon$--terms with appropriate coefficients at each order. We have not been able to find a
generic prescription for the construction of such terms from the components of the conserved
currents of the complete $AdS_4\times CP^3$ superstring, and the brute force computation becomes
technically more and more involved with each new order in fermions. So to simplify the analysis we
shall turn to the consideration of a simpler $AdS_4$ sub--sector of the theory in which the problem
of the construction of the Lax connection can be completely solved at least in a particular
kappa--symmetry gauge.

\section{String in $\mathcal N=2$ $AdS_4$ superspace}\label{superspace}
\setcounter{equation}0

As has been shown in \cite{Gomis:2008jt} the structure of the $AdS_4\times CP^3$ superstring action
and equations of motion allows one to consistently truncate this theory to a model describing a
string propagating in a four--dimensional superbackground with eight fermionic directions
parameterized by $\upsilon=(1-\mathcal P_{24})\Theta$. The bosonic subspace of this superbackground
is $AdS_4$ but it does not preserve any supersymmetry\footnote{A somewhat analogous
non--supersymmetric $AdS_4$ vacuum was found in a matter--coupled ${\mathcal N}=2,$ $D=4$
supergravity in
\cite{Cassani:2009ck}.}.
This model is obtained by putting to zero the 24 supersymmetric fermionic fields $\vartheta=\mathcal
P_{24}\,\Theta=0$ and restricting the string to move entirely in $AdS_4$ (\emph{i.e.} the $CP^3$
embedding coordinates are worldsheet constants). It is, therefore, not described by the supercoset
sigma--model of
\cite{Arutyunov:2008if,Stefanski:2008ik,Fre:2008qc,Bonelli:2008us,D'Auria:2008cw}. Lacking
supersymmetry this model is also \emph{not} the $\frac{OSp(2|4)}{SO(1,3)\times SO(2)}$ supercoset
sigma--model  \cite{Gomis:2008jt}. Nevertheless, it possesses the four--parameter kappa--symmetry
in addition to the purely bosonic isometry $SO(2,3)$ of $AdS_4$ and the $SO(2)$ symmetry rotating
the two $D=4$ Majorana fermions. So it is somewhat surprising that this model turns out to be integrable, and the
integrability of its fermionic sector is not at all related to target space supersymmetry which is
lacking.

Let us consider this model in more detail. It is convenient to represent the eight--component
spinors $\upsilon=(1-\mathcal P_{24})\Theta$ as four--component Majorana spinors in $AdS_4$,
$\upsilon^{\alpha i}$ $(\alpha=1,2,3,4)$, carrying the internal $SO(2)$ index $i=1,2$. This $SO(2)$
is a relic of the $U(1)$ gauge symmetry associated with the RR one--form field of $D=10$ type IIA
supergravity. The Green--Schwarz action for the superstring moving in this $AdS_4$ superspace has
the following form
\cite{Gomis:2008jt}
\begin{equation}\label{cordaA}
S = -\frac{1}{4\pi\alpha'}\,\int d^2\xi\, \sqrt {-h}\, h^{IJ}\, {\cal E}_I{}^{\hat a} {\cal
E}_J{}^{\hat b}
\eta_{\hat a\hat b} -\frac{1}{2\pi\alpha'}\,\int  B_2\,,
\end{equation}
where the vector supervielbeins ${\cal E}^{\hat a}=dx^{\hat m}\,{\cal E}_{\hat m}{}^{\hat
a}+d\upsilon^{\alpha i}{\cal E}_{\alpha i}{}^{\hat a}$ along the $AdS_4$ directions of the target
superspace are
\be\label{simplA}
\begin{aligned}
{\mathcal E}^{\hat a}(x,\upsilon) &= e^{{1\over3}\phi(\upsilon)}\,\left(e^{\hat
b}(x)+4i\upsilon\gamma^{\hat b}\,{{\sinh^2{{\mathcal M}/ 2}}\over{\mathcal
M}^2}\,D\upsilon\right)\Lambda_{\hat b}{}^{\hat a}(\upsilon)
\\
&{}
\hskip+1cm +e^{-{1\over3}\phi(\upsilon)}\,\frac{4R}{kl_p}\upsilon\,\ve\gamma^5\,{{\sinh^2{{\mathcal
M}/2}}\over{\mathcal M}^2}\,D\upsilon\, V{}^{\hat a}(\upsilon)\,,
\end{aligned}
\ee
and the NS--NS superform $B_2$ is expressed through components of its field strength $H_3=dB_2$ as
follows
\be\label{B2}
B_2=\int_0^1\,dt\,i_\upsilon H_3(x,t\upsilon)\,\,.
\ee
\be\label{H3}
\begin{aligned}
H_3&=dB_2=-\frac{1}{3!}{\mathcal E}^{\hat c}{\mathcal E}^{\hat b}{\mathcal E}^{\hat
a}(\frac{6}{kl_p}e^{-\phi}\ve_{\hat a\hat b\hat c\hat d}V{}^{\hat d}) +{\mathcal E}^{\hat
a}{\mathcal E}^{\beta j}{\mathcal E}^{\alpha i}(\gamma_{\hat
a}\gamma^{5})_{{\alpha\beta}}\,\varepsilon_{ij} -{\mathcal E}^{\hat b}{\mathcal E}^{\hat
a}{\mathcal E}^{\alpha i}(\gamma_{\hat a\hat b}\gamma^5\,\varepsilon\lambda)_{\alpha i}\,,
\end{aligned}
\ee
where ${\mathcal E}^{\alpha i}(x,\upsilon)$ are the fermionic supervielbeins
\be\label{simplB}
\begin{aligned}
{\mathcal E}^{\alpha i}(x,\upsilon) &= e^{{1\over6}\phi(\upsilon)}\,\left({{\sinh{\mathcal
M}}\over{\mathcal M}}\,D\upsilon\right)^{\beta j}\,S_{\beta j}{}^{\alpha i}\,(\upsilon)
-ie^{\phi(\upsilon)}{\mathcal A}_1(x,\upsilon)\,(\gamma^5\varepsilon\lambda(\upsilon))^{\alpha i}
\end{aligned}
\ee
and ${\mathcal A}_1(x,\upsilon)$ is a relic of the type IIA RR one--form
\be\label{simplA1}
\begin{aligned}
{\mathcal A}_1(x,\upsilon) &=\frac{R}{kl_p}\,e^{-{4\over3}\phi(\upsilon)}\,\left[
\left(e^{\hat a}(x)+4i\upsilon\gamma^{\hat
a}\,{{\sinh^2{{\mathcal M}/2}}\over{\mathcal M}^2}\,D\upsilon\right)V_{\hat a}(\upsilon)
-4\upsilon\,\ve\gamma^5\,{{\sinh^2{{\mathcal M}/2}}\over{\mathcal M}^2}\,D\upsilon\,\Phi(\upsilon)
\right]\,.
\end{aligned}
\ee
Note that $\mathcal A_1$ is zero when $\upsilon=0$.

The $AdS_4$ covariant derivative $D$ is defined as
\bee\label{D}
D\upsilon=\left(\nabla+\frac{i}{R}e^{\hat a}(x)\,\gamma^5\gamma_{\hat
a}\right)\upsilon=\left(d-\frac{1}{4}\omega^{\hat a
\hat b}(x)\,\gamma_{\hat a\hat b}+\frac{i}{R}e^{\hat a}(x)\,\gamma^5\gamma_{\hat a}\right)\upsilon \,
\eee
and $\gamma^{\hat a},\,\gamma^5$ are the four--dimensional gamma--matrices in the Majorana
representation.

 The dilaton superfield $\phi(\upsilon)$, which depends only on
the eight fermionic coordinates, has the following form in terms of the quantities $V{}^{\hat
a}(\upsilon)$ and $\Phi(\upsilon)$
\be\label{dilaton1}
e^{{2\over 3}\phi(\upsilon)}={R\over{kl_p}}\,\sqrt{\Phi^2+V{}^{\hat a}\,V{}^{\hat b}\,\eta_{\hat
a\hat b}}\,.
\ee
The value of the dilaton at $\upsilon=0$ is
\begin{equation}
e^{\frac{2}{3}\phi(\upsilon)}|_{\upsilon=0}=e^{\frac{2}{3}\phi_0}=\frac{R}{kl_p}\,
\end{equation}
($l_p$ is the Plank's length and $k$ corresponds to the Chern--Simons level in the ABJM model). The
fermionic field $\lambda^{\alpha i}(\upsilon)$ describes the non--zero components of the dilatino
superfield which is related to the dilaton superfield by the equation
\cite{Howe:2004ib}
\be\label{dilatino1}
\lambda_{\alpha i}=-\frac{i}{3}D_{\alpha i}\,\phi(\upsilon)\,.
\ee

The new objects appearing in these expressions, $\mathcal M$, $\Lambda_{\hat a}{}^{\hat b}$, $\Phi$, $V{}^{\hat a}$ and $S_{\alpha i}{}^{\beta j}$, are functions of $\ups$ and their explicit forms are
given in Appendix B. Contracted spinor indices have been suppressed,
\emph{e.g.} $(\ups\ve\gamma^5)_{\alpha i}=\ups^{\beta j}\ve_{ji}\gamma^5_{\beta\alpha}$, where
$\varepsilon_{ij}=-\varepsilon_{ji}$, $\varepsilon_{12}=1$ is the $SO(2)$ invariant tensor.

As we have already noted, in the $AdS_4$ superspace under consideration all supersymmetries are
broken and it only has the bosonic $AdS_4$ isometry $SO(2,3)$. The superstring action
\eqref{cordaA} is thus invariant under the $SO(2,3)$ variations of the coordinates
\be\label{varixv}
\delta x^{\hat m}\,e_{\hat m}{}^{\hat a}(x)=K^{\hat a}(x)=K^{\hat m}(x)\,e_{\hat m}{}^{\hat a}\,,\qquad
\delta\upsilon=\frac{1}{4}(K^{\hat m}\,\omega_{\hat m}{}^{\hat a\hat b}(x)
-\nabla^{\hat a}\,K^{\hat b})\,\gamma_{\hat a\hat b}\,\upsilon.
\ee
 The associated conserved $SO(2,3)$ current has the following form
\be\label{J23}
J^I=\sqrt{-h}\,h^{IJ}\,\mathcal E_{J}{}^{\hat a}\,(i_{\delta x}\,\mathcal E{}^{\hat
b}+i_{\delta\upsilon }\,\mathcal E{}^{\hat b})\,\eta_{\hat a\hat b}-\varepsilon^{IJ}\,(i_{\delta
x}\,B_2+i_{\delta\upsilon}\,B_2)_J\,.
\ee
Due to the complicated form of the supervielbein and $B_2$, the explicit dependence of this current
on $\upsilon$ is still a bit too involved to try to construct a Lax connection. So we shall further
simplify things by gauge fixing kappa--symmetry in a way considered in
\cite{Grassi:2009yj}.

\section{Gauge fixed superstring action in $AdS_4$ superspace}
\setcounter{equation}0

Let us choose the $AdS_4$ metric in the conformally flat form
\be\label{ads4metric21}
ds^2_{_{AdS_4}}={1\over u^2}(dx^a\eta_{ab}dx^b+\frac{R_{CP^3}^2}{4}\,du^2)\,,
\qquad u=\left(R_{CP^3}\over r\right)^2\,,
\ee
where $x^a$ $(a=0,1,2)$ are the coordinates of the $D=3$ Minkowski boundary and $u$ (or $r$) is the
$AdS_4$ radial coordinate. If the components of the $AdS_4$ vielbein associated with the metric
(\ref{ads4metric21}) are chosen to be\footnote{Note that the vielbeins $e^a$ and $e^3$ appearing in
eq. (\ref{ad4v}) correspond to the $AdS_4$ metric of the $D=11$ $AdS_4\times S^7$ solution
characterized by the radius R which is related to the $CP^3$ radius in the string frame as follows
$R_{CP^3} = e^{\frac{1}{3}\phi_0}R =\left(\frac{R^3}{kl_p}\right)^{1/2}\,$. These bosonic vielbeins
appear in our explicit expressions for the $AdS_4$ supergeometry.}
\be\label{ad4v}
e^{\frac{\phi_0}{3}}\,e^a={r^2\over R_{CP^3}^2}\,dx^a=u^{-1}\,dx^a\,,\qquad
e^{\frac{\phi_0}{3}}\,e^3=\frac{R_{CP^3}}{r}\,dr=-\frac{R_{CP^3}}{2u}\,du,
\ee
the components of the $SO(1,3)$ spin connection are
\be\label{eaoa31}
\omega^{a3}=-\frac{2}{R}\,e^a\,,
\ee
and
\be\label{oab}
\omega^{ab}=0\,,
\ee
where the index 3 stands for the 3rd (radial) direction in $AdS_4$.

The following kappa--symmetry gauge fixing condition on $\upsilon$ drastically simplifies the form
of the superstring action
\be\label{kgf}
\upsilon=\frac{1}{2}(1+\gamma^{012})\,\upsilon\,,\qquad
\gamma^{012}\equiv \gamma\,,
\ee
where $\gamma^{012}$ is the product of the gamma matrices along the 3d Minkowski boundary slice of
$AdS_4$.

In this gauge the supervielbeins  take the following simple form \cite{Grassi:2009yj}
\be
\begin{aligned}
{\mathcal E}^a(x,\upsilon) &=(\frac{R}{kl_p})^{1/2}
(e^a(x)+i\ups\gamma^aD\,\ups)\,(1-\frac{1}{R^2}(\ups\ups)^2)\,,
\\
\\
{\mathcal E}^3(x,\upsilon)
&=(\frac{R}{kl_p})^{1/2}e^3(x)\,(1-\frac{3}{R^2}(\ups\ups)^2)\,,
\\
\end{aligned}
\ee
and the covariant derivative becomes
\be
D\upsilon=\left(d-\frac{1}{R}e^3(x)-\frac{1}{4}\omega^{ab}(x)\,\gamma_{ab}\right)\upsilon\,.
\ee
Actually, the $SO(1,2)$ Lorentz connection $\omega^{ab}$ is zero when the $AdS_4$ supervielbeins
are taken in the form \eqref{ad4v}.

The NS-NS two form becomes
\bee\label{simple+B}
B_2&=& -\frac{i}{kl_p}
\Big[
(e^b+i\upsilon\gamma^bD\upsilon)\,
(e^a+i\upsilon\gamma^aD\upsilon)\,
\ups\gamma^c\ve\ups\,\varepsilon_{abc}
 -R\,e^3\,\upsilon\varepsilon D\upsilon
\Big]\,.
\eee

The kappa--symmetry gauge--fixed superstring action reduces to
\begin{eqnarray}\label{cordaB}
S &=&-\frac{1}{4\pi\alpha'}\,\frac{R}{k l_p}
\int\,d^2\xi\,\sqrt{-h}\,h^{IJ}\left[
e_I{}^3 e_J{}^3\,(1-\frac{6}{R^2}(\ups\ups)^2)
\right.
\nonumber \\
&&{}
\hspace{4cm}
+\left.(e_I{}^a+i\upsilon\gamma^aD_I\upsilon)\,
(e_J{}^b+i\upsilon\gamma^bD_{J}\upsilon)\,\eta_{ab}\,
(1-\frac{2}{R^2}(\ups\ups)^2)\right]
\nonumber\\
\\
 &+&\frac{1}{2\pi\alpha'}\frac{i}{kl_p}\int
\Big[
(e^b+i\upsilon\gamma^bD\upsilon)\,
(e^a+i\upsilon\gamma^aD\upsilon)\,
\ups\gamma_{ab}\ve\ups
 -R\,e^3\,\upsilon\varepsilon D\upsilon
\Big].\nonumber
\end{eqnarray}
This action is slightly more complicated than the action for the
$AdS_5\times S^5$ superstring in the analogous kappa--symmetry gauge
\cite{Kallosh:1998ji}. The latter contains fermions only up to the fourth order.

In this kappa--symmetry gauge, the conserved $SO(2,3)$ current \eqref{J23} has the following
explicit form
\bee\label{J23xplic}
J&=&\sqrt{-h}\,h^{IJ}\,e_J{}^{\hat a}\,K_{\hat
a}+i\upsilon(\sqrt{-h}\,h^{IJ}-i\varepsilon^{IJ}\,\gamma^5\varepsilon)\gamma^{\hat
a}\nabla_J\upsilon\,K_{\hat a}\nonumber
\\&-&\frac{i}{4}\,\upsilon(\sqrt{-h}\,h^{IJ}+i\varepsilon^{IJ}\,\gamma^5\varepsilon)\gamma_{\hat
a}{}^{\hat b\hat c}\,\upsilon\,e_J{}^{\hat a}\nabla_{\hat b}\,K_{\hat c}\\
&-&\!\!\!\!\sqrt{-h}\,h^{IJ}\,\left[\frac{(\upsilon\upsilon)^2}{2R^2}(e_J{}^a\,K_a+12\,e_J{}^3\,K_3)
+\frac{3\,(\upsilon\upsilon)^2}{8R}e_J{}^a\,(\nabla_3\,K_a-\nabla_a\,K_3)
-\frac{\upsilon\upsilon}{4}\varepsilon^{abc}\,\upsilon\gamma_a\nabla_J\upsilon\,\nabla_bK_c\right]\nonumber\\
&-&\frac{3}{2R}\,\varepsilon^{IJ}\,\upsilon\gamma_a\nabla_J\upsilon\,\upsilon\gamma^{ab}\varepsilon\upsilon\,K_b
-\frac{1}{8}\varepsilon^{IJ}\,\upsilon\gamma_a\nabla_J\upsilon\,\upsilon\gamma^{ab}\varepsilon\upsilon(\nabla_3\,K_b-\nabla_b\,K_3)\,,\nonumber
\eee
where remember that $\varepsilon$ without indices implies $\varepsilon^{ij}=-\varepsilon^{ji}$ with
$i,j=1,2$ labeling the two $D=4$ Majorana fermions $\upsilon^{\alpha i}$. The first two lines in
\eqref{J23xplic} are the same as in the quadratic current \eqref{JB} reduced to $AdS_4$ and with
$\vartheta=0$. The third and the fourth line are quartic in $\upsilon$ and its derivative.

 The problem of the construction of the Lax connection thus becomes
more treatable, but we would like to simplify things even further.

\subsection{Worldsheet T--dual action for the $AdS_4$ superstring}
Upon a T--duality transformation on the worldsheet
\cite{Grassi:2009yj}, similar to that described in
\cite{Kallosh:1998ji} \footnote{Note that in contrast to the $AdS_5\times S^5$ superstring
where this bosonic T--duality can be accompanied by a fermionic one
\cite{Ricci:2007eq,Berkovits:2008ic,Beisert:2008iq} which brings the superstring action to itself
 but in a different kappa--symmetry gauge, in the $AdS_4\times CP^3$ case the fermionic T--duality
is not possible \cite{Adam:2009kt,Grassi:2009yj}, at least in the same fashion and in application
to the broken supersymmetry fermions $\upsilon$. For an alternative suggestion to perform bosonic
and fermionic worldsheet T--duality of the $AdS_4\times CP^3$ supercoset model see
\cite{Bargheer:2010hn,Henn:2010ps,Huang:2010qy} and for problems with its realization see \cite{Adam:2010hh}.}, the action (\ref{cordaB}) takes an even simpler
form
\begin{eqnarray}\label{cordaBT}
S &=&-\frac{1}{4\pi\alpha'}\,\frac{R}{k l_p}
\int\,d^2\xi\,\sqrt{-h}\,h^{IJ}
\left({\tilde e}_I{}^a\,{\tilde e}_J{}^b\,\eta_{ab}+e_I{}^3 e_J{}^3\right)\,(1-\frac{6}{R^2}(\ups\ups)^2)
\nonumber\\
\\
 &-&\frac{1}{2\pi\alpha'}\frac{iR}{kl_p}\int\Big(
e^3\,\upsilon\varepsilon D\upsilon +{\tilde e}^a\,\upsilon\gamma_aD\upsilon -\frac{1}{R}\,{\tilde
e}^a\,{\tilde e^b}\,\ups\gamma_{ab}\ve\ups\Big),\nonumber
\end{eqnarray}
where $e^3(r)$ and $\tilde e^{a}(\tilde x,r)$ are the vielbeins of the \emph{dual} $AdS_4$ space.
The dual vielbeins $\tilde e^{a}(\tilde x,r)$ along the Minkowski directions  are related to the
initial quantities as follows (see
\cite{Grassi:2009yj} for more details)
\be\label{trcurrent}
\partial_I\,({r^2\over R^2}\,P^I_a)=0\, \qquad\Rightarrow
\qquad P^I_a=\frac{R^2}{r^2}\,\varepsilon^{IJ}\partial_J\,{\tilde
x}_a\equiv \varepsilon^{IJ}\,{\tilde e}_{Ja}
\ee
where
\be\label{P}
P^I_a=-\sqrt{-h}\,(1-\frac{2}{R^2}(\upsilon\upsilon)^2)\,
\Big(h^{IJ}\eta_{ab}
+\frac{2i}{R\sqrt{-h}}\,\varepsilon^{IJ}\,\ups\gamma_{ab}\ve\ups\Big)
\,(e_J{}^b+i\upsilon\gamma^bD_J\upsilon)\,.
\ee
The quantities (\ref{P}) (up to a rescaling) are  the conserved currents of the $d=3$ translation
part of the $SO(2,3)$ isometries
\be\label{trans}
\delta\, {\tilde x}^a=c^a, \qquad \delta\,r=\delta\,\upsilon=0\,.
\ee
The action \eqref{cordaBT} can be cast in the manifestly $SO(1,3)$ covariant form
\begin{eqnarray}\label{cordaBT1}
S =-\frac{1}{2\pi\alpha'}\,\frac{R}{k l_p}
\int\,d^2\xi\,\left(\frac{1}{2}\sqrt{-h}\,h^{IJ}
{\tilde e}_I{}^{\hat a}\,{\tilde  e}_J{}^{\hat b}\,\eta_{\hat a\hat b}
\,(1-\frac{6}{R^2}(\ups\ups)^2)
+i\varepsilon^{IJ}\,{\tilde  e}_I{}^{\hat a}\,\upsilon(1-\Gamma_{11})\,\gamma_{\hat
a}{\nabla}_J\upsilon\right)
\,,\nonumber\\
\end{eqnarray}
where $\Gamma_{11}$ stands for $(1-\mathcal P_{24})\gamma^5\,\gamma^7(1-\mathcal P_{24})\equiv
i\gamma^5\,\varepsilon$ which indicates its origin from $D=10$ and
$\nabla=d-\frac{1}{4}\,\tilde
\omega^{\hat a\hat b}\,\gamma_{\hat a \hat b}$.

The bosonic and fermionic equations of motion which follow from
\eqref{cordaBT1} are, respectively,
\be\label{bosonic}
{\nabla}_I\left(\sqrt{-h}\,h^{IJ}\,{\tilde e}_J{}^{\hat a}\,(1-\frac{6}{R^2}(\ups\ups)^2)
+i\varepsilon^{IJ}\,\upsilon(1-\Gamma_{11})\,\gamma^{\hat a}\nabla_J\upsilon\right)
-\frac{2i}{R^2}\,\varepsilon^{IJ}\,{\tilde e}_I{}^{\hat b}\,{\tilde e}_J{}^{\hat c}\,\upsilon(1-\Gamma_{11})\,\gamma^{\hat a}{}_{\hat b\hat
c}\upsilon=0
\ee
and
\be\label{fermionic}
\frac{i}{2}\,(1+\gamma)(1-\Gamma^{11})\,\varepsilon^{IJ}\,{\tilde e}_I{}^{\hat
a}\,\gamma_{\hat a}\,{\nabla}_J\,\upsilon-\frac{6}{R^2}\,\upsilon\,
(\upsilon\upsilon)\,\sqrt{-h}\,h^{IJ} \,{\tilde e}_I{}^{\hat a}\,{\tilde e}_J{}^{\hat b}\,\eta_{\hat
a\hat b}=0\,,
\ee
The fermionic equation can also be rewritten in the following form
\bee\label{fermionic2}
i\sqrt{-h}h^{IJ}{\tilde e}_J{}^{\hat a}\,(1-\Gamma_{11})\,\gamma_{\hat a}\,\nabla_{I}\upsilon
-\frac{6}{R^2}\,\varepsilon^{IJ}{\tilde e}_I{}^{\hat a}\,{\tilde e}_J{}^{\hat
b}\,(1-\Gamma_{11})\gamma_{\hat a\hat b}\upsilon\,(\upsilon\upsilon)=0\,, &
\eee

The conserved current of the $SO(2,3)$ isometry is
\be\label{J}
J^I=\left(\sqrt{-h}\,h^{IJ}\,{\tilde e}_J{}^{\hat
a}\,(1-\frac{6}{R^2}(\ups\ups)^2)+i\varepsilon^{IJ}\,\upsilon(1-\Gamma_{11})\,\gamma^{\hat
a}{\nabla}_J\upsilon\right)\,K_{\hat a}+\frac{i}{4}\,\varepsilon^{IJ}\,{\tilde
e}_J{}^{\hat a}\,\upsilon\,(1-\Gamma_{11})\,\gamma_{\hat a}{}^{\hat b\hat
c}\,\upsilon\,{\nabla}_{\hat b}\,K_{\hat c}\,.
\ee

\subsection{The Lax connection}\label{D4L}
 As in the quadratic approximation of Section \ref{laxquad},
 we construct an $SO(2,3)$--valued zero--curvature Lax connection $\mathcal L$
\be\label{R=0}
{\cal R}=d\mathcal L-\mathcal L\mathcal L=0\,\qquad \Longrightarrow \qquad
\varepsilon^{IJ}(\partial_I\,\mathcal L_J+\mathcal L_I\,\mathcal L_J)=0\,
\ee
using the pieces of the conserved current
\eqref{J} which enter the Lax connection with arbitrary coefficients. The problem has a non--trivial solution if
the zero--curvature condition allows for expressing the coefficients  in terms of a single spectral
parameter. In the case under consideration the zero--curvature Lax connection has the following
form
\bee\label{LAX}
\mathcal L_I&=&\alpha_1 \,{\tilde e}_I{}^{\hat a}\,K_{\hat a}+\alpha_2\,\frac{\varepsilon_{IJ}}{-h}\,J^J
+\frac{\alpha_2^2}{\sqrt{-h}}\,F_I+\alpha_1\alpha_2\,\frac{\varepsilon_{IJ}}{-h}\,F^J\nonumber\\
&&-\frac{\alpha^2_2}{4R^2}\,
\upsilon\,(1-\Gamma_{11})\,\gamma_{\hat a}\,\nabla_I\,\upsilon\,\upsilon\,(1-\Gamma_{11})\,\gamma^{\hat a\hat b\hat
c}\,\upsilon\,K_{\hat b}\,K_{\hat c}\\
&& +\frac{3\alpha^2_2}{2R^2}(\ups\ups)^2\,{\tilde e}_I{}^{\hat a}\,K_{\hat
a}+\frac{3\alpha_2(\alpha_1+2)}{8}\,\partial_I\left(\frac{(\upsilon\upsilon)^2}{\sqrt{-G}}\,\varepsilon^{JK}\,
{\tilde e}_J{}^{\hat a}\,{\tilde e}_K{}^{\hat b}\,K_{\hat a}K_{\hat b}\right)\,,\nonumber
\eee
where $G=\det({\tilde e_I{}^{\hat a}\,\tilde e_J{}^{\hat b}\,\eta_{\hat a\hat b}})$,
\be\label{F2}
F^I=\frac{i}{4}\,\varepsilon^{IJ}\,{\tilde e}_J{}^{\hat a}\,\upsilon\,(1-\Gamma_{11})\,\gamma_{\hat
a}{}^{\hat b\hat c}\,\upsilon\,{\nabla}_{\hat b}\,K_{\hat
c}\,=\frac{i}{2}\,\varepsilon^{IJ}\,{\tilde e}_J{}^{\hat
a}\,\upsilon\,(1-\Gamma_{11})\,\gamma_{\hat a}{}^{\hat b\hat c}\,\upsilon\,K_{\hat b}\,K_{\hat c}
\ee
and (as in Section \ref{laxquad})
 \be\label{alpha2}
 \alpha^2_2=\alpha^2_1+2\alpha_1.
 \ee
So the Lax connection contains one independent (spectral) parameter $\alpha_1=\frac{2z^2}{1-z^2}$.

 To check that \eqref{LAX} has zero curvature one should use the string equations of motion
\eqref{bosonic}--
\eqref{fermionic2}, the Killing vector relations (Appendix D) and the Fierz identities (Appendix C).

Note that in the kappa--symmetry gauge under consideration the Lax connection is of the fourth
order in fermions (as is the action \eqref{cordaBT1} and the conserved current \eqref{J}).

Applying the inverse duality transformation \eqref{trcurrent} to \eqref{LAX} one gets the Lax
connection for the original model \eqref{cordaB} which is non--local in the coordinates $x^a$ of
the Minkowski boundary of the $AdS_4$ space \eqref{ads4metric21}, the non--local quantities being
the Killing vectors $K_A(\tilde x(x),r)$ and their derivatives $\nabla_A\,K_B(\tilde x(
x),r)=[K_A,K_B]$ expressed in terms of the original $AdS_4$ coordinates. With some more technical
effort, it should be possible to construct an alternative local Lax connection of the model
\eqref{cordaB} directly from the conserved current \eqref{J23xplic}. We leave this exercise for
future consideration.

\section{Conclusion and Discussion}\label{cd}

In this paper we have constructed the full Lax connection for the $AdS_4$ sub--sector of the
$AdS_4\times CP^3$ superstring with eight `broken supersymmetry' fermionic modes which is not
described by the supercoset sigma--model. Because of the technical complexity of the problem, the
construction has been carried out for the kappa--symmetry gauge fixed and worldsheet T--dualized
action of the theory. For a generic (semi)classical configuration of the $AdS_4\times CP^3$
superstring with 32 fermionic fields (which are not subject to a kappa--symmetry gauge fixing) we
have constructed the Lax connections up to the second order in the fermionic fields. These results
provide a direct evidence for the classical integrability of the complete $AdS_4\times CP^3$
superstring theory.

It would be useful, though, to find a procedure for the construction of a Lax connection of the
complete theory to all orders in the thirty two fermions. A hint at a possible method to achive
this goal may come from the construction of Lax connections in two--dimensional supersymmetric
$O(N)$ and $CP^N$ sigma--models. When these sigma--models are formulated in components of
corresponding $d=2$ supermultiplets, a prescription for constructing the Lax connection was proposed
in
\cite{Curtright:1979am} which, as we have already mentioned, has prompted the techniques used in
this paper. A more systematic way of constructing the Lax connections for these supersymmetric
sigma--models is in the framework of their worldsheet superfield description which allows one to
operate with a corresponding Cartan superform or a conserved super--current in the worldsheet
superspace rather than with their components \cite{Chau:1981gj,Ivanov:1987yv,Saleem:2006tv}.

One can try to develop similar methods for studying the classical integrability of Green--Schwarz
superstrings in the framework of the superembedding approach (see
\cite{Bandos:1995zw,Howe:1997wf,Sorokin:1999jx} for review and references). The superembedding
description of superparticles, superstrings and superbranes is based on the fact that the
worldsheet kappa--symmetry is a somewhat weird realization of the conventional extended worldsheet
supersymmetry \cite{Sorokin:1989zi}. The dynamics of $p$--branes is described by an embedding of a
worldsheet supersurface into a target superspace subject to a certain superembedding condition. The
embedding super--coordinates $X^M$ and $\Theta^{\underline\alpha}$ of a superstring in this
formulation are therefore worldsheet superfields, as in the case of two--dimensional supersymmetric
sigma--models and the Ramond--Neveu--Schwarz strings. A difference is that in the latter the
component (bosonic and fermionic) worldsheet fields are in the same supermultiplet, while in the
superembedding approach $X$ and $\Theta$ are (a priori) in different supermultiplets and
corresponding superfields. However, these superfields are related to each other by the
superembedding constraint which (at least in some cases) can be solved in terms of a single
`prepotential'. The superembedding formulation is intrinsically related to super--twistors
\cite{Sorokin:1989zi,Volkov:1988vf,Gumenchuk:1990db} and pure--spinors \cite{Tonin:1991ii,Matone:2002ft}. It has proved to be
extremely useful \emph{ e.g.} for the derivation of the M5--brane equations of motion
\cite{Howe:1996yn,Howe:1997fb} and for making progress in the covariant description of multiple coincident
branes \cite{Howe:2005jz,Howe:2006rv,Howe:2007eb,Bandos:2009yp,Bandos:2009gk}.

To construct a Lax connection for a superstring in the superembedding approach one should first
derive a conserved worldsheet supercurrent associated with the superisometry of the supergravity
background under consideration and then try to use it in combination with a spectral parameter for
building the worldsheet superfield Lax connection. The expansion of this Lax connection in
worldsheet superfield components should then reproduce the form of Lax connections considered in
this paper to all orders in fermions. We hope to address this problem in the near future.

Other possible applications and development of the results of this paper can be the generalization
to the complete $AdS_4\times CP^3$ superstring of the algebraic curve constructed in
\cite{Gromov:2008bz} and the study of the integrability of type IIB superstrings compactified on
$AdS_3\times S^3\times S^3\times S^1$ and on $AdS_3\times S^3\times T^4$ (with 16 preserved
supersymmetries) in those sectors which are not described by corresponding supercoset sigma--models
(see
\cite{Babichenko:2009dk,Zarembo:2010yz} and references therein). An even more interesting case is type II
superstrings in an $AdS_2\times S^2\times T^6$ superbackground which preserves only eight
supersymmetries and is related to the near horizon geometry of $D=4$ black holes
\cite{Klebanov:1996mh}. In this case 16 independent kappa--symmetries are not enough to eliminate
24 `broken supersymmetry' fermions and hence the $\frac{PSU(1,1|2)}{S0(1,1)\times U(1)}$ supercoset
sigma--model
\cite{Zhou:1999sm} cannot be regarded as a kappa--gauge fixed description of this theory.

One may also look for other examples of integrable superstrings in superbackgrounds with less or no
supersymmetry, whose purely bosonic sub--sector is integrable. As we have seen in Section
\ref{superspace}, the superstring in the $\mathcal N=2$ $AdS_4$ superspace is integrable in spite of
the fact that all the eight supersymmetries are broken. If we did not know that this
non--supersymmetric model is a truncation of the $AdS_4\times CP^3$ superstring, we would wonder
what might be the reason for its integrability. An obvious further example to check for
integrability is the superstring in the $AdS_4\times CP^3$ background with all supersymmetries
broken. This superbackground is obtained from the 24--supersymmetric solution by changing the sign
of the $F_2$ flux \cite{Nilsson:1984bj}. We have not been able to construct a zero--curvature Lax
connection for this case using the technique developed in this paper. So it still remains to be
understood what is the deep reason for the integrability of the $AdS_4\times CP^3$ superstring in
the fermionic sub--sector corresponding to the broken supersymmetries. Does this indicate that the
superstring in $AdS_4\times CP^3$ remembers that it is obtained by the dimensional reduction of the
maximally supersymmetric $AdS_4\times S^7$ superbackground of $D=11$ supergravity
\cite{Nilsson:1984bj,Sorokin:1984ca,Sorokin:1985ap}?
\\
\\
{\bf Acknowledgements}\\
The authors are grateful to Alessandra Cagnazzo for collaboration at early stages of this project
and to G. Arutyunov, I. Bandos, N. Berkovits, Jaume Gomis, E. Ivanov, Z. Popowicz, A. Tseytlin, and
K. Zarembo for useful comments and discussion. This work was partially supported by the INFN
Special Initiative TV12. D.S. was also partially supported by an Excellence Grant of Fondazione
Cariparo (Padova) and the grant FIS2008-1980 of the Spanish Ministry of Science and Innovation.
D.S. is grateful to the Organziers of the 2010 Simons Workshops ``Superstrings in Ramond--Ramond
backgrounds" and ``Supersymmetry, Strings and Gauge Theory" for their hospitality and financial
support during his stay at Stony Brook University where part of this work was done.

\def\thesection{}
\def\theequation{A.\arabic{equation}}\label{A}
\section{Appendix A. Main notation and conventions}
\setcounter{equation}0

The convention for the ten--dimensional metric is the `almost plus' signature $(-,+,\cdots,+)$.
Generically, the tangent space vector indices are labeled by letters from the beginning of the
Latin alphabet,  while  letters from the middle of the Latin alphabet stand for curved (world)
indices. The spinor indices are labeled by Greek letters.

\def\thesubsection{A.1}
\subsection{$AdS_4$ space}

$AdS_4$ is parametrized by the coordinates $x^{\hat m}$ and its vielbeins are $e^{\hat a}=dx^{\hat
m}\,e_{\hat m}{}^{\hat a}(x)$, ${\hat m}=0,1,2,3;$ ${\hat a}=0,1,2,3$. The $D=4$ gamma--matrices
satisfy:
\be\label{gammaa}
\{\gamma^{\hat a},\gamma^{\hat b}\}=2\,\eta^{\hat a\hat b}\,,
\qquad \eta^{\hat a\hat b}={\rm diag}\,(-,+,+,+)\,,
\ee
\be\label{gamma5}
\gamma^5=i\gamma^0\,\gamma^1\,\gamma^2\,\gamma^3, \qquad
\gamma^5\,\gamma^5=1\,.
\ee
The charge conjugation matrix $C$ is antisymmetric, the matrices $(\gamma^{\hat
a})_{\alpha\beta}\equiv (C\,\gamma^{\hat a})_{\alpha\beta}$ and $(\gamma^{\hat a\hat
b})_{\alpha\beta}\equiv(C\,\gamma^{\hat a\hat b})_{\alpha\beta}$ are symmetric and
$\gamma^5_{\alpha\beta}\equiv (C\gamma^5)_{\alpha\beta}$ is antisymmetric, with
$\alpha,\beta=1,2,3,4$ being the indices of a 4--dimensional spinor representation of $SO(1,3)$ or
$SO(2,3)$.

The $AdS_4$ curvature is
\be\label{ads4c}
R_{\hat a\hat b\hat c}{}^{\hat d}=\frac{8}{R^2}\,\eta_{\hat c[\hat a}\,\delta_{\hat b]}^{\hat
d}\,,\qquad R^{\hat a\hat b}=-\frac{4}{R^2}\,e^{\hat a}\,e^{\hat b}\,,
\ee
where $\frac{R}{2}$ is the $AdS_4$ radius.

\def\thesubsection{A.2}
\subsection{$CP^3$ space}

$CP^3$ is parametrized by the coordinates $y^{m'}$ and its vielbeins are
$e^{a'}=dy^{m'}e_{m'}{}^{a'}(y)$, ${m'}=1,\cdots,6;$ ${a'}=1,\cdots,6$. The $D=6$ gamma--matrices
satisfy:
\be\label{gammaa'}
\{\gamma^{a'},\gamma^{b'}\}=2\,\delta^{{a'}{b'}}\,,\qquad \delta^{a'b'}={\rm
diag}\,(+,+,+,+,+,+)\,,
\ee
\be\label{gamma7}
\gamma^7={i\over{6!}}\,\varepsilon_{\,a_1'a_2'a_3'a_4'a_5'a_6'}\,\gamma^{a_1'}\cdots \gamma^{a_6'} \qquad
\gamma^7\,\gamma^7=1\,.
\ee
The charge conjugation matrix $C'$ is symmetric and the matrices
$(\gamma^{a'})_{\alpha'\beta'}\equiv (C\,\gamma^{a'})_{\alpha'\beta'}$ and
$(\gamma^{a'b'})_{\alpha'\beta'}\equiv(C'\,\gamma^{a'b'})_{\alpha'\beta'}$ are antisymmetric, with
$\alpha',\beta'=1,\cdots,8$ being the indices of an 8--dimensional spinor representation of
$SO(6)$.

The $CP^3$ curvature is
\be\label{cp3c}
R_{a'b'c'}{}^{d'}=-\frac{2}{R^2}\,\Big(\delta_{c'[a'}\,\delta^{d'}_{b']}-J_{c'[a'}\,J_{b']}{}^{d'}+J_{a'b'}\,J_{c'}{}^{d'}\Big).
\ee

\def\thesubsection{A.3}
\subsection{The $D=10$ gamma--matrices $\Gamma^A$}
\bee\label{Gamma10}
&\{\Gamma^A,\,\Gamma^B\}=2\eta^{AB},\qquad
\Gamma^{A}=(\Gamma^{\hat a},\,\Gamma^{a'})\,,\nonumber\\
&\\
&\Gamma^{\hat a}=\gamma^{\hat a}\,\otimes\,{\bf 1},\qquad
\Gamma^{a'}=\gamma^5\,\otimes\,\gamma^{a'},\qquad
\Gamma^{11}=\gamma^5\,\otimes\,\gamma^7,\qquad \hat a=0,1,2,3;\quad
a'=1,\cdots,6\,. \nonumber
\eee
The charge conjugation matrix is ${\mathcal C}=C\otimes C'$.

The fermionic variables $\Theta^{\underline\alpha}$ of IIA supergravity carrying 32--component
spinor indices of $Spin(1,9)$, in the $AdS_4\times CP^3$ background and for the above choice of the
$D=10$ gamma--matrices, naturally split into 4--dimensional $Spin(1,3)$ indices and 8--dimensional
spinor indices of $Spin(6)$, i.e. $\Theta^{\underline\alpha}=\Theta^{\alpha\alpha'}$
($\alpha=1,2,3,4$; $\alpha'=1,\cdots,8$).

\def\thesubsection{A.4}
\subsection{$OSp(6|4)$ superalgebra}\label{ospalgebra}
The bosonic part of the $OSp(6|4)$ algebra is generated by translations and Lorentz-transformations which split into $AdS_4$ and $CP^3$ parts as
$P_A=(P_{\hat a},P_{a'})$ and $M_{AB}=(M_{\hat a\hat b},M_{a'b'})$ respectively. These satisfy the commutation relations
\be\label{sp411}
[P_A,P_B]=-\frac{1}{2}R_{AB}{}^{CD}M_{CD},\qquad [M_{AB},P_C]=\eta_{AC}\,P_B-\eta_{BC}\,P_A\,,
\ee
\be
[M_{AB},M_{CD}]=\eta_{AC}\,M_{BD}+\eta_{BD}\,M_{AC}-\eta_{BC}\,M_{AD}-\eta_{AD}\,M_{BC}\,,
\ee
where the curvature $R_{AB}{}^{CD}=(R_{\hat a\hat b}{}^{\hat c\hat d},R_{a'b'}{}^{c'd'})$, and the $AdS_4$ and $CP^3$
curvature are given in (\ref{ads4c}) and (\ref{cp3c}) respectively. The fermionic part of the
algebra consists of 24 supersymmetry generators which can be described by 32--component Majorana
spinor generators subject to the projection $Q_{\underline\alpha}=(\mathcal
P_{24}\,Q)_{\underline\alpha}$ (see eq. \eqref{p6}). Their commutation relations are as follows
\bee\label{osp64}
&[P_A,\,Q]=\frac{i}{R}\,Q{\gamma^5\Gamma_A}\mathcal P_{24}\,,\quad [M_{AB},\,Q]=-{1\over
2}\,Q\,\Gamma_{AB}\mathcal P_{24}\,,\quad \\
\nonumber\\
&\{Q,Q\}=2i\,(\mathcal P_{24}\Gamma^A\mathcal P_{24})\,P_A+\frac{R}{4}(\mathcal
P_{24}\gamma^5\Gamma^{AB}\mathcal P_{24})R_{AB}{}^{CD} M_{CD}\,,\nonumber
\eee
where $\gamma_5=i\Gamma^0\Gamma^1\Gamma^2\Gamma^3$. Note that the splitting of the fermionic generators $Q$ into $Q_1$ and $Q_3$
by the $Z_4$--grading of $OSp(6|4)$ is simply achieved by splitting the $D=10$ Majorana spinor $Q$
into the left-- and right Majorana--Weyl spinors
\be\label{q1q3}
Q_1=\frac{1}{2}\,Q\,(1-\Gamma_{11})\,,\qquad Q_3=\frac{1}{2}\,Q\,(1+\Gamma_{11})\,.
\ee

\def\thesubsection{B}
\def\theequation{B.\arabic{equation}}
\section{Appendix B. Quantities appearing in the definition of the
$AdS_4\times CP^3$ superspace of Section \ref{superspace}}\label{AB}
\setcounter{equation}0

\be\label{M}
R\,({\mathcal M}^2)^{\alpha i}{}_{\beta j}= 4(\ve\upsilon)^{\alpha i}(\ups\ve\gamma^5)_{\beta j}
-2(\gamma^5\gamma^{\hat a}\upsilon)^{\alpha i}(\ups\gamma_{\hat a})_{\beta j} -(\gamma^{\hat a\hat
b}\upsilon)^{\alpha i}(\ups\gamma_{\hat a\hat b}\gamma^5)_{\beta j}\,,
\ee
\be
\begin{aligned}
\Lambda_{\hat a}{}^{\hat b}&=
\delta_{\hat a}{}^{\hat b}-\frac{R^2}{k^2l_p^2}\,\cdot\,
\frac{e^{-\frac{2}{3}\phi}}{e^{\frac{2}{3}\phi}
+{R\over{kl_p}}\,\Phi}\,{V_{\hat a}}\,V_{}^{\hat b}\,,
\\
\\
S_{\beta j}{}^{\alpha i}&=
\frac{e^{-\frac{1}{3}\phi}}{\sqrt2}\left[(1-\mathcal P_{24})\,\left(\sqrt{e^{\frac{2}{3}\phi}
+{R\over{kl_p}}\,\Phi}\,{\bf 1}-{R\over{kl_p}}\,
\frac{V{}^{\hat a}\,\Gamma_{\hat a}\Gamma_{11}}{\sqrt{e^{\frac{2}{3}\phi}
+{R\over{kl_p}}\,\Phi}}
\,\right)(1-\mathcal P_{24})\right]_{\beta j}{}^{\alpha i}
\end{aligned}
\ee

\be\label{phiE7}\begin{aligned}
V{}^{\hat a}(\upsilon)&=-\frac{8i}{R}\,\upsilon\gamma^{\hat a}\,{{\sinh^2{{\mathcal M}/
2}}\over{\mathcal M}^2}\,\varepsilon\,{\upsilon}\,,
\\
\Phi(\upsilon)&= 1+\frac{8}{R}\,\upsilon\,\ve\gamma^5\,{{\sinh^2{{\mathcal
M}/2}}\over{\mathcal M}^2}\,\ve\upsilon\,.
\end{aligned}
\ee
Let us emphasise that the $SO(2)$ indices $i,j=1,2$ are raised and lowered with the unit matrices
$\delta^{ij}$ and $\delta_{ij}$ so that there is actually no difference between the upper and the
lower $SO(2)$ indices, $\varepsilon_{ij}=-\varepsilon_{ji}$, $\varepsilon^{ij}=-\varepsilon^{ji}$
and $\varepsilon^{12}=\varepsilon_{12}=1$.

\def\theequation{C.\arabic{equation}}
\section{Appendix C. Identities for the kappa-projected fermions}\label{C}
\setcounter{equation}0

When the fermionic variables $\upsilon^{\alpha i}$ are subject to the constraint (\ref{kgf}), the
following identities hold.
\be\label{i1}
\ups^i\gamma^5\ups^j=\ups^i\gamma^3\ups^j=0\,,\qquad \ups^{\alpha i}\ups^{\beta j}\delta_{ij}
=-\frac{1}{4}((1+\gamma)C^{-1})^{\alpha\beta}\ups\ups\,,
\ee
where $\gamma=\gamma^{012}$ and $\ups\ups=\delta_{ij}\ups^{\alpha i}C_{\alpha\beta}\ups^{\beta j}$.

Another useful relation is ($\ve^{012}=-\ve_{012}=1$)
\be\label{gg}
\ups\gamma_{ab}d\ups=\pm\ve_{abc}\ups\gamma^cd\ups\,.
\ee

Using eqs. (\ref{i1}) and (\ref{gg}) we find that
\be
\ups\ve\gamma^a\ups\,\ups\ve\gamma_b\ups=\delta_b^a(\ups\ups)^2\,,
\qquad \ups\ve\gamma^{ac}\ups\,\ups\ve\gamma_{cb}\ups=2\delta_b^a(\ups\ups)^2\,,
\ee
and
\bee
({\mathcal M}^2\ve\ups)^{\alpha i}=0\,.
\eee
 A similar computation shows that
\be
\ups\ve\gamma^5{\mathcal M}^2=0.
\ee
It is also true in general (i.e. without fixing $\kappa$--symmetry) that
\be
{\mathcal M}^2\ups=0\,,\qquad \ups\gamma^5{\mathcal M}^2=0.
\ee
Using the above identities we find that for $\upsilon$ satisfying (\ref{kgf})
\be\label{M2d}
\mathcal M^2D\upsilon
=\frac{6i}{R^2}(e^a+\frac{R}{2}\omega^{a3})(\gamma_a\ups)\,\ups\ups
\ee
which results in
\be\label{vmdv}
4\upsilon\gamma^a \frac{\sinh^2(\mathcal M/2)}{\mathcal M^2}D\upsilon=
\upsilon\gamma^a(1+\frac{1}{12}\mathcal M^2)D\upsilon
=\upsilon\gamma^a\,(d-\frac{1}{4}\omega^{bc}\gamma_{bc})\upsilon
+\frac{i}{2R^2}(e^a+\frac{R}{2}\omega^{a3})(\ups\ups)^2\,,
\ee
where $e^a$, $e^3$ $\omega^{bc}$ and $\omega^{a3}$ are $AdS_4$ vielbeins and connection defined in
eqs. (\ref{ad4v})--\eqref{oab} and the matrix ${\mathcal M}^2$ is defined in eq. (\ref{M}).

\noindent
We also find that
\be
4\ups\ve\gamma^5{{\sinh^2{{\mathcal M}/2}}\over{\mathcal M}^2}D\ups
=\ups\ve\gamma^5D\ups=\frac{i}{R}(e^a+\frac{R}{2}\omega^{a3})\ups\ve\gamma_a\ups\,.
\ee
Other $D=4$ covariant Fierz identities $(\hat a=(a,3))$ used in the construction of the Lax
connection in Section \ref{D4L} are
\bee\label{Fierz1}
\varepsilon^{IJ}\,{\nabla}_I\upsilon\,(1-\Gamma_{11})\,\gamma_{\hat a}\,\nabla_J\,\upsilon\,\upsilon\,(1-\Gamma_{11})\,\gamma^{\hat a\hat b\hat
c}\,\upsilon-2\,\varepsilon^{IJ}\,\upsilon\,(1-\Gamma_{11})\,\gamma^{\hat
b}\,\nabla_I\,\upsilon\,\upsilon\,(1-\Gamma_{11})\,\gamma^{\hat
c}\,\nabla_J\,\upsilon\,= \nonumber\\
\\
 =\frac{1}{2}\,\varepsilon^{IJ}\nabla_I\left(\upsilon\,(1-\Gamma_{11})\,\gamma_{\hat a}\,\nabla_J\,\upsilon\,\upsilon\,(1-\Gamma_{11})\,\gamma^{\hat a\hat b\hat
c}\,\upsilon\right)-\frac{2}{R^2}\,
\varepsilon^{IJ}\,e_I{}^{\hat b}\,e_J{}^{\hat c}\,
(\upsilon\upsilon)^2\,,\nonumber
\eee
\be\label{ident}
\upsilon(1-\Gamma^{11})\gamma_{\hat a\hat c\hat d}\upsilon\,\upsilon(1-\Gamma^{11})\gamma^{\hat b\hat c\hat d}\upsilon
=-6\delta^{\hat b}_{\hat a}\,(\upsilon\upsilon)^2=6\upsilon(1-\Gamma^{11})\gamma_{\hat
a}\gamma_5\upsilon\,\upsilon(1-\Gamma^{11})\gamma^{\hat b}\gamma_5\upsilon\,.
\ee
where $\Gamma_{11}$ stands for $(1-\mathcal P_{24})\gamma^5\,\gamma^7(1-\mathcal P_{24})\equiv
i\gamma^5\,\varepsilon$ which indicates its origin from $D=10$.

\def\theequation{D.\arabic{equation}}
\section{Appendix D. Basic relations for the Killing vectors on symmetric spaces $G/H$}\label{AD}
\setcounter{equation}0
Let $K_M(X)$ or $K_A(X)=e_A{}^M(X)\,K_M(X)$ be the Killing vectors of a $D$--dimensional symmetric
space $G/H$, where $M$ are world indices and $A$ are tangent space indices. The Killing vectors
$K_M(X)$ take values in the algebra of the isometry group $G$ and the one--forms $K=dX^M\,K_M$
satisfy the Maurer--Cartan equations
\bee\label{K}
dK=-2K\wedge K,\qquad dK\wedge K= K\wedge dK=-2K\wedge K \wedge K.
\eee
The following relations also hold
\bee\label{K1}
&[\nabla_A,\nabla_B]K_C=-R_{ABC}{}^D\,K_D\,,\qquad \nabla_AK_B=[K_A,K_B],\qquad& \\
&\nabla_A\nabla_BK_C=[\nabla_AK_B,\,K_C]+[K_B,\nabla_AK_C]=
[\nabla_AK_B,\,K_C]-[\nabla_AK_C,\,K_B]=-2R_{A[BC]}{}^DK_D,\,\,\,\,\,\,\,~&\\
&\,[\nabla_AK_B,\,K_C]=[[K_A,\,K_B],\,K_C]=-R_{ABC}{}^D\,K_D,&\\
&\Big[[K_A,K_B],[K_C,K_D]\Big]=R_{AB[C}{}^F\,[K_{D]},K_F]-R_{CD[A}{}^F\,[K_{B]},K_F]\,,&
\eee
where $R_{ABC}{}^D$ is the curvature of the symmetric space $G/H$.

For instance, for the $AdS_4$ Killing vectors we have
\bee\label{K2}
&[\nabla_{\hat a},\nabla_{\hat b}]K_{\hat c}=-R_{\hat a\hat b\hat c}{}^{\hat d}\,K_{\hat
d}\,,\qquad R_{\hat a\hat b\hat c}{}^{\hat d}=\frac{8}{R^2}\,\eta_{\hat c[\hat a}\,\delta_{\hat
b]}^{\hat d},\,\qquad
R^{\hat a\hat b}=-\frac{4}{R^2}\,e^{\hat a}\,e^{\hat b}\,,&\\
&\nabla_{\hat a}K_{\hat b}=[K_{\hat a},K_{\hat b}],\qquad& \\
&\nabla_{\hat a}\nabla_{\hat b}K_{\hat c}=[\nabla_{\hat a}K_{\hat b},\,K_{\hat c}]+
[K_{\hat b},\nabla_{\hat a}K_{\hat c}]= [\nabla_{\hat a}K_{\hat b},\,K_{\hat c}]
-[\nabla_{\hat a}K_{\hat c},\,K_{\hat b}]=\frac{8}{R^2}\,\eta_{{\hat a}[\hat b}\,K_{\hat c]},&\\
&\,[\nabla_{\hat a}K_{\hat b},\,K_{\hat c}]=[[K_{\hat a},\,K_{\hat b}],\,K_{\hat c}]
=-\frac{8}{R^2}\,\eta_{{\hat c}[\hat a}\,K_{\hat b]}&\\
&\Big[[K_{\hat a},K_{\hat b}],[K_{\hat c},K_{\hat d}]\Big]=-\frac{16}{R^2}\,(K_{[\hat
c}\,\eta_{\hat d][\hat a}\,K_{\hat b]}-K_{[\hat a}\,\eta_{\hat b][\hat c}\,K_{\hat d]}).&
\eee


\providecommand{\href}[2]{#2}\begingroup\raggedright\endgroup

\end{document}